# A High Quality/Low Computational Cost Technique for Block Matching Motion Estimation

S. López, G.M. Callicó, J.F. López and R. Sarmiento
Research Institute for Applied Microelectronics (IUMA)
Department of Electronic Engineering and Control (DIEA)
University of Las Palmas de Gran Canaria, E-35017, Spain
seblopez@iuma.ulpgc.es

**Abstract**

*Motion estimation is the most critical process in video coding systems. First of all, it has a definitive impact on the rate-distortion performance given by the video encoder. Secondly, it is the most computationally intensive process within the encoding loop. For these reasons, the design of high-performance low-cost motion estimators is a crucial task in the video compression field. An adaptive cost block matching (ACBM) motion estimation technique is presented in this paper, featuring an excellent tradeoff between the quality of the reconstructed video sequences and the computational effort. Simulation results demonstrate that the ACBM algorithm achieves a slight better rate-distortion performance than the one given by the well-known full search algorithm block matching algorithm with reductions of up to 95% in the computational load.*

## 1 Introduction

Video compression systems are based on exploiting the spatial and temporal redundancies present in a digital video sequence. Following this hybrid strategy, high compression ratios at reasonable video quality levels are obtained. In order to achieve a good compromise between these two features, motion estimation techniques play a key role in the process of removing temporal redundancies between consecutive frames, and hence, it is considered the most critical part in high performance video encoders.

Although many motion estimation methods have been proposed, block matching (BM) algorithms are the most popular ones because of their simplicity, robustness and ease of implementation [1]. These algorithms are based on dividing a current frame into a block of $N \times M$ pixels (typically 16×16 pixels), called reference block, which is then compared with blocks of identical size, called candidate blocks, within a search area of size $(N+2p) \times (M+2p)$ contained in the previous frame, where $p$ is the maximum allowed displacement. The displacement between the coordinates of the block in the current frame and the best matched block in the search area gives as result the motion vector, as it is shown in Fig.1

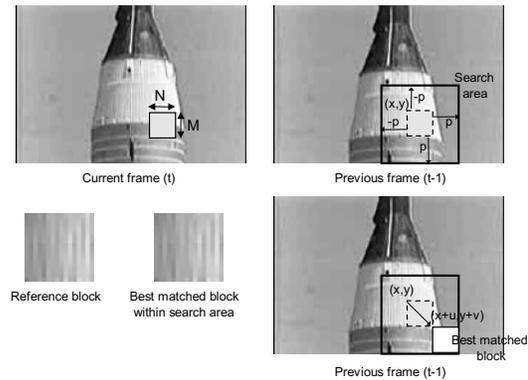

**Fig. 1: Block matching technique**

The full search block matching (FSBM) algorithm is the most popular block matching technique, because it obtains an excellent image quality at the decoder side with a low control overhead [2]. However, this optimal performance is achieved at the expenses of evaluating every possible candidate motion vector within the search area, representing a prohibitive cost solution for real time low power applications. In order to overcome this problem, several fast block matching motion estimation algorithms have been proposed in the recent literature. These algorithms can be classified into two groups depending on the employed strategy: on one hand, those based on reducing the number of search points [3, 4, 5]; on the other hand, the ones based on reducing the number of pixels used for block matching [6, 7, 8]. Predictive block matching (PBM) algorithms are included within the first strategy. These algorithms have attracted the interest of the research community, improving the performance of previously reported fast algorithms by exploiting the spatio-temporal correlation that exhibit motion fields in real video sequences [9, 10]. Due to the intrinsic nature of these algorithms, an extremely low computational cost and a high correlated motion vectors field are guaranteed, although acceptable quality levels are restricted to very low bit rates in slow motion video sequences sampled at high frame rates.

A novel hybrid solution, named adaptive cost block matching (ACBM) algorithm, that combines the benefits of PBM and FSBM algorithms is proposed in this paper. Among other features, the ACBM presents independence of the processed video sequence while providing a much higher decoded image quality than PBM with a considerable reduced computational cost related to FSBM. In this sense, our algorithm represents a highly flexible strategy in order to control, depending on the potential application, the weight given to video quality or computational load, guaranteeing a good compromise between these two features.

The rest of this paper is organized as follows. In Section 2, a detailed description of the PBM and FSBM algorithms is presented, showing their advantages and disadvantages. Section 3 reports a description of the ACBM algorithm while in Section 4 the simulation results obtained are highlighted. Finally, in Section 5, the conclusions of this work are outlined.

## 2 Algorithms comparison

### 2.1 General scenario

It is well known that the performance of a generic hybrid video encoder can be enhanced by using Lagrangian optimization techniques. The application of these techniques to the motion estimation process results in the minimization of the following cost function:

$$J(mv_x, mv_y) = D(mv_x, mv_y) + \lambda \cdot R(mv_x, mv_y)$$

where $(mv_x, mv_y)$ represents the candidate motion vector, $\lambda$ is the Lagrange operator, proportional to the quantization step $Q_p$, $R(mv_x, mv_y)$ represents the total number of bits needed to transmit the candidate motion vector and finally, $D(mv_x, mv_y)$, is the distortion for this vector or matching error, measured as the sum of absolute differences (*SAD*), defined as:

$$SAD(mv_x, mv_y) = \sum_{i=0}^{N-1} \sum_{j=0}^{M-1} | p_t(i,j) - p_{t-1}(i+mv_x, j+mv_y) |$$

where $N \times M$ is the block size and $p_t$, $p_{t-1}$ represent the luminance pixel values of the current and reference blocks respectively. It is important to note that the $J(mv_x, mv_y)$ cost function represents an excellent metric in order to compare the performance given by several motion estimation algorithms, in the sense that the best motion vector for a macroblock is the one that minimizes this function.

### 2.2 PBM algorithms

These algorithms are based on the hypothesis that, in real digital video sequences, the motion field varies slowly in the spatial and temporal directions. Under this assumption, it is proper to think that, previously computed motion vectors in a spatio-temporal neighbourhood should be very similar to the motion vector to be computed for the reference block. PBM algorithms operate in three steps [9,10]. First of all, a set of candidate predictors is chosen from the spatio-temporal neighborhood of the current block. The motion vectors that compose this neighborhood are shown in Fig.2, being the temporal and spatial neighbors the motion vectors in the previous and current frame respectively.

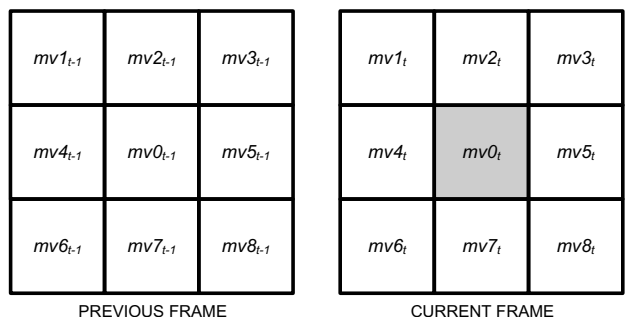

**Fig. 2: Spatio-temporal neighbourhood**

In order to select a set of predictors for the computation of the motion vector of the reference block (shadowed), the motion vectors denoted as $mv5_t$ .. $mv8_t$ cannot be selected, as they have not been computed yet. The second step in PBM algorithms is to select the candidate with lowest SAD. Finally, a refinement around the best predictor is performed in order to obtain a motion vector that reduces the prediction error as much as possible. Normally, the refinement step is performed in a half pixel grid, allowing the achievement of motion vectors with this precision.

Following this three-steps procedure, PBM algorithms only evaluate a reduced set of motion vectors, achieving a very low computational cost. In addition, PBM algorithms achieve a very smooth and coherent motion vector field that minimizes the $R(mv_x, mv_y)$ term in the $J(mv_x, mv_y)$ cost function, since motion vectors are differentially encoded in actual video coding standards. Unfortunately, the rate-distortion performance given by these algorithms is sequence dependent, in the sense that they tend to fail when dealing with high textured and/or sharp motion sequences. In these conditions, the PBM algorithms get easily trapped into the local minimum, which causes a much higher matching error when compared with the FSBM algorithm.

### 2.3 The FSBM algorithm

The FSBM algorithm evaluates all the possible candidates within a previously established search area, selecting the position with minimal SAD at the end of the process. This means that, in order to obtain an integer pixel motion vector for one block, the FSBM algorithm evaluates $(2p+1)^2$ positions. If a half pixel precision motion vector is required, the FSBM considers 8 additional half pixel candidates around the position pointed by the integer pixel motion vector.

As the FSBM evaluates all the positions inside the search area, it greatly minimizes the matching error $D(mv_x,mv_y)$ and hence, the cost function $J(mv_x,mv_y)$. This fact makes FSBM algorithm to be considered as a near optimal performance solution, in terms of high peak signal to noise ratio (PSNR), providing a much better rate-distortion characteristic when compared with the PBM algorithms. However, the FSBM algorithm exhibits a prohibitive computational cost and suffers from poor motion vector allocation. This last drawback is derived from the fact that this algorithm simply finds a motion vector that minimizes a displaced frame difference error, and does not consider reproducing the real motion of the scene. For this reason, the motion field given by the FSBM algorithm is normally incoherent, and the number of bits $R(mv_x,mv_y)$ needed to transmit the computed motion vectors increases with respect to the PBM algorithms.

## 3 Proposed motion estimation algorithm

The motion estimation technique presented in this paper combines the benefits of FSBM and PBM into a novel smart strategy. Our algorithm is based on the idea of applying the FSBM algorithm exclusively on those image blocks where it is completely necessary, in order to maintain the rate-distortion performance within certain quality limits with a low computational effort.

### 3.1 Preliminary studies

The experimental setup shown in Figure 3 has been designed in order to detect the situations in which the use of the FSBM algorithm can be avoided [11]. The methodology is as follows: a ten frames sequence is generated by using an original reference frame, introducing nine different global motion vectors perfectly known. After that, the FSBM algorithm is applied over this sequence, and the results are compared on a block-by-block basis with the original motion vectors previously introduced, detecting true and false motion vectors.

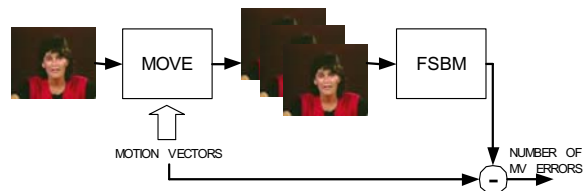

**Figure 3. Experimental setup**

In addition, two new parameters are also obtained in this procedure. The first one is the Intra Sum of Absolute Differences (*Intra_SAD*) parameter, defined as:

$$Intra\_SAD = \sum_{i=0}^{N-1}\sum_{j=0}^{M-1} | p_t(i,j) - \mu |$$

being computed for each block of the current frame with µ representing the average value of the pixels in the whole block. This parameter is very helpful in order to distinguish high textured blocks, characterized by high *Intra_SAD* values.

The second introduced parameter is called Sum of Absolute Differences Deviation (*SAD_deviation*), being defined as:

$$SAD\_deviation = \sum_{u,v} |SAD(u,v) - SAD\_min|$$

where *SAD(u,v)* denotes the *SAD* from the evaluated candidate at position *(u,v)*, and *SAD_min* represents the minimum *SAD* obtained from all the evaluated positions.

This experimental setup is applied on numerous video sequences obtaining several data like the ones presented in Fig. 4. This figure shows six graphs where the *Intra_sad* and the *SAD_deviation* for all the 16×16 pixels blocks (graph dots) of the whole sequence are represented in horizontal and vertical axes respectively, with the blocks where a true motion vector was obtained appearing in the upper left graph (error=0 graph), and in the rest of them, the blocks with motion vector errors (error=1,2,3,4 and error≥5 graphs). From this evaluation, two main conclusions have been extracted:

- High textured blocks usually have associated true, and hence coherent, motion vectors.
- These blocks present high *SAD_deviation* and *SAD_min* values

These two statements reveal that, for high textured blocks, it is not recommendable to apply a PBM algorithm as $J(mv_x,mv_y)$ dramatically increases if the minimum *SAD* position pointed by the FSBM algorithm is not selected. On the other hand, for low textured blocks, only a minimal decrease in the matching error is obtained if the FSBM algorithm is applied, at the expenses of very appreciable increments in the computational cost and in the number of bits, $R(mv_x,mv_y)$, needed to transmit a big amount of uncorrelated motion vectors. For this reason, the application of the FSBM algorithm is not justified when dealing with high textured blocks.

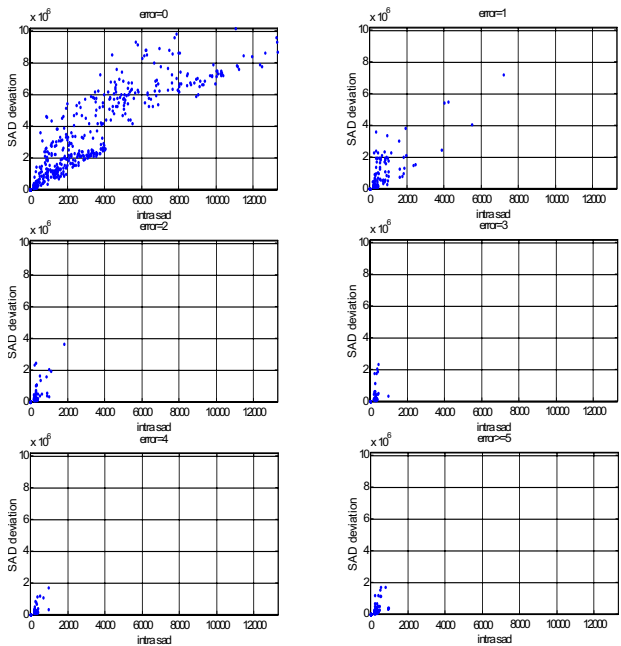

**Figure 4. Experimental setup results**

### 3.2 The ACBM algorithm

The ACBM algorithm proposed in this paper has been designed considering not only the experimental results mentioned before, but also the fact that PBM algorithms, due to the nature of the cost function $J(mv_x,mv_y)$, perform their best when high quantization steps are used.

In order to obtain a motion vector, the ACBM algorithm firstly computes the *Intra_SAD* that corresponds to the reference block. Once this parameter has been computed for the current block, the PBM algorithm described in [9] is applied and due to the reasons explained in the previous section, the following condition is evaluated:

$$Intra\_SAD + SAD\_PBM < \alpha + \beta \cdot Q_p^2$$

where *SAD_PBM* is the *SAD* associated to the motion vector found by the PBM algorithm while $\alpha$ and $\beta$ are fixed parameters related to the desired levels of quality and computational cost. If this condition holds true, the motion estimation process is finished and it is not necessary to run the FSBM algorithm. Even if the *Intra_SAD* value forces this condition not to be met, it is possible to avoid the use of FSBM algorithm as long as the PBM algorithm finds a motion vector that exhibits a minimal or close to minimal *SAD*. This condition is modelled by:

$$SAD\_PBM < \gamma \cdot Intra\_SAD$$

where $\gamma$ controls if the *SAD* obtained with the PBM algorithm is considered low enough for the high textured block under analysis. If none of these two conditions are true, then the block is identified as a critical one and the FSBM algorithm must be applied in order to avoid an important degradation in the quality of the reconstructed block.

It is important to remark that the ACBM algorithm represents a flexible motion estimation solution in the sense that the computational cost, and hence the video quality, can be easily controlled by modifying the values of $\alpha$, $\beta$ and $\gamma$ parameters. In that sense, the algorithm can be adjusted in order to avoid the use of the FSBM algorithm for all the image blocks. However, the key feature of the presented algorithm is the excellent compromise achieved between computational cost and reconstructed image quality as it will be shown in the next section.

## 4 Simulation results

In order to evaluate the performance of the ACBM algorithm, several simulations were carried out for different video sequences using an H.263 encoder with half pixel precision [12]. For this purpose, a considerable amount of QCIF (176×144 pixels) and CIF sequences (352×288 pixels) sampled at 30, 15 and 10 frames per second were selected. With all these sequences, several simulations were performed with different α, β and γ values, comparing the rate-distortion performance of the ACBM algorithm with the ones given by the PBM algorithm and the FSBM algorithm with *p*=15.

After these exhaustive tests, and as a particular case of study to test the goodness of the ACBM algorithm, the values of α, β and γ were fixed in order to obtain similar quality levels to the ones obtained with the FSBM algorithm, being the values of 1000, 8 and ¼ the best options respectively. Figure 5 shows the results obtained with the designed algorithm on QCIF sequences sampled at 30 frames per second, demonstrating an important improvement in the rate-distortion characteristic when compared with the one given by the PBM algorithm. The same feature is observed in Figure 6, where QCIF sequences were also selected, this time sampled at 10 frames per second. For both cases, it is also shown that a slight better rate-distortion performance than the one obtained with the FSBM algorithm is achieved for all the selected sequences, with independence of texture, type and amount of movement in the scene or frame rate of the video sequence.

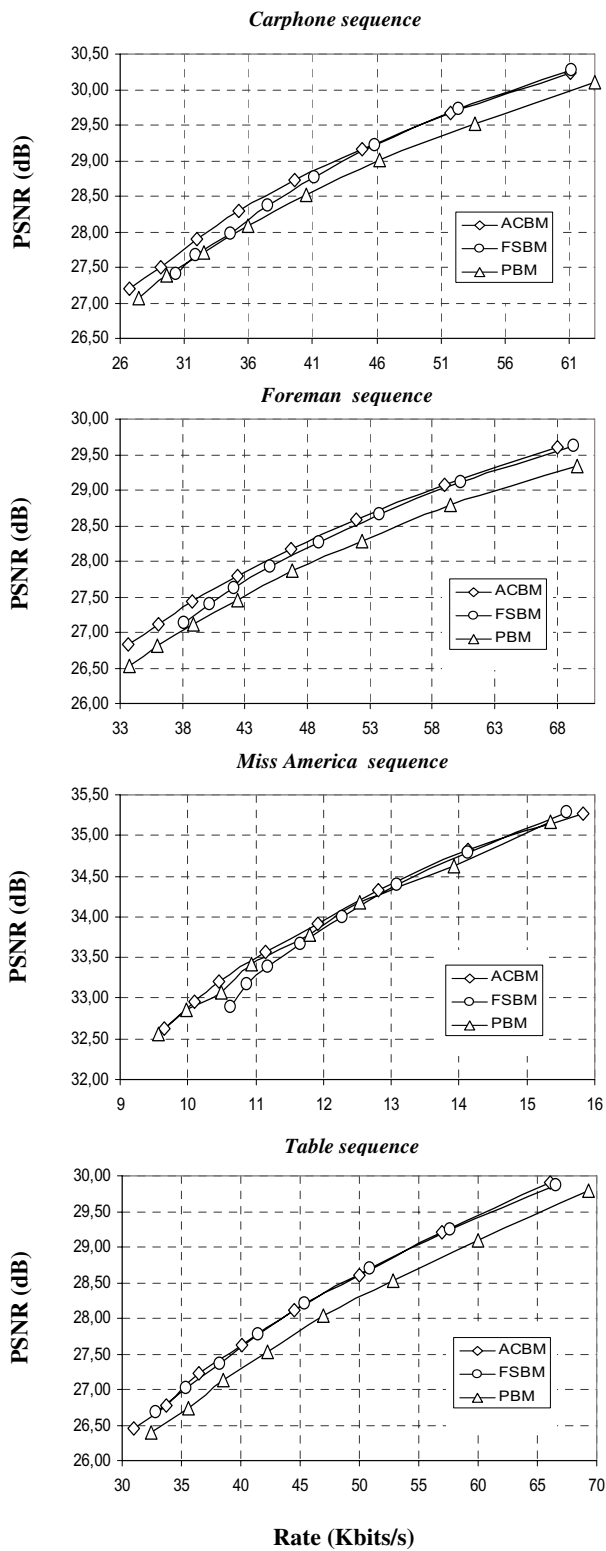

**Figure 5. Simulation results with QCIF@30 fps**

It is important to note that the difference between the rate-distortion performance obtained with the PBM and the ACBM algorithms becomes larger as the frame rate decreases. This trend is completely logical because for low frame rates sequences, the motion field does not vary slowly in the temporal direction, as it is supposed by the PBM algorithms. Due to this reason, PBM algorithms perform their best at high frame rates (typically 30 frames per second) and the ACBM algorithm exhibits a lower computational cost, as the number of blocks in which the use of FSBM is avoided increases. These experimental results are highlighted in Table 1 where the computational complexity is presented in terms of the average number of candidate positions searched per macroblock for the ACBM algorithm. The benefits of this algorithm are even more appreciable when compared to FSBM algorithm, for which 969 candidate positions are evaluated. From Table 1 it is also observed that, for homogeneous and smooth motion sequences (Miss America sequence) the ACBM technique presents the lowest computational cost, while for high textured and abrupt motion sequences (Foreman sequence), it is the highest. This result is directly linked with the rate-distortion curves shown in Figs. 5 and 6 where the difference between the performance given by the PBM and the FSBM algorithm strongly depends on the selected video sequence.

| Sequence | Carphone | | Foreman | | Miss America | | Table | |
|---|---|---|---|---|---|---|---|---|
| $Q_p$ | 30 fps | 10 fps | 30 fps | 10 fps | 30 fps | 10 fps | 30 fps | 10 fps |
| 30 | 274 | 313 | 313 | 381 | 42 | 61 | 235 | 265 |
| 28 | 323 | 352 | 381 | 458 | 61 | 71 | 274 | 294 |
| 26 | 371 | 410 | 441 | 536 | 90 | 100 | 313 | 333 |
| 24 | 410 | 458 | 507 | 594 | 100 | 119 | 352 | 371 |
| 22 | 458 | 487 | 555 | 642 | 119 | 138 | 390 | 420 |
| 20 | 468 | 507 | 584 | 671 | 129 | 148 | 420 | 449 |
| 18 | 478 | 526 | 604 | 700 | 138 | 158 | 458 | 487 |
| 16 | 497 | 555 | 613 | 720 | 168 | 187 | 507 | 536 |

**Table 1. Computational complexity**

## 5 Conclusions

A novel adaptive cost block matching (ACBM) motion estimation algorithm has been presented in this paper, showing a full range of promising applications in the multimedia processing scenario. The introduction of key parameters, together with the election of their values, permits the achievement of a good trade-off between image quality and low complexity costs. The most remarkable results show that for a similar PSNR than the one exhibited by FSBM, a considerable reduction in the complexity load is obtained. Furthermore, our algorithm is self-adapted to different frame rates, and hence, it is suitable for variable bandwidth channel conditions. Innovative architectural solutions are right now under development in our research group, based on sharing common resources to FSBM and PBM architectures applied to portable multimedia devices.

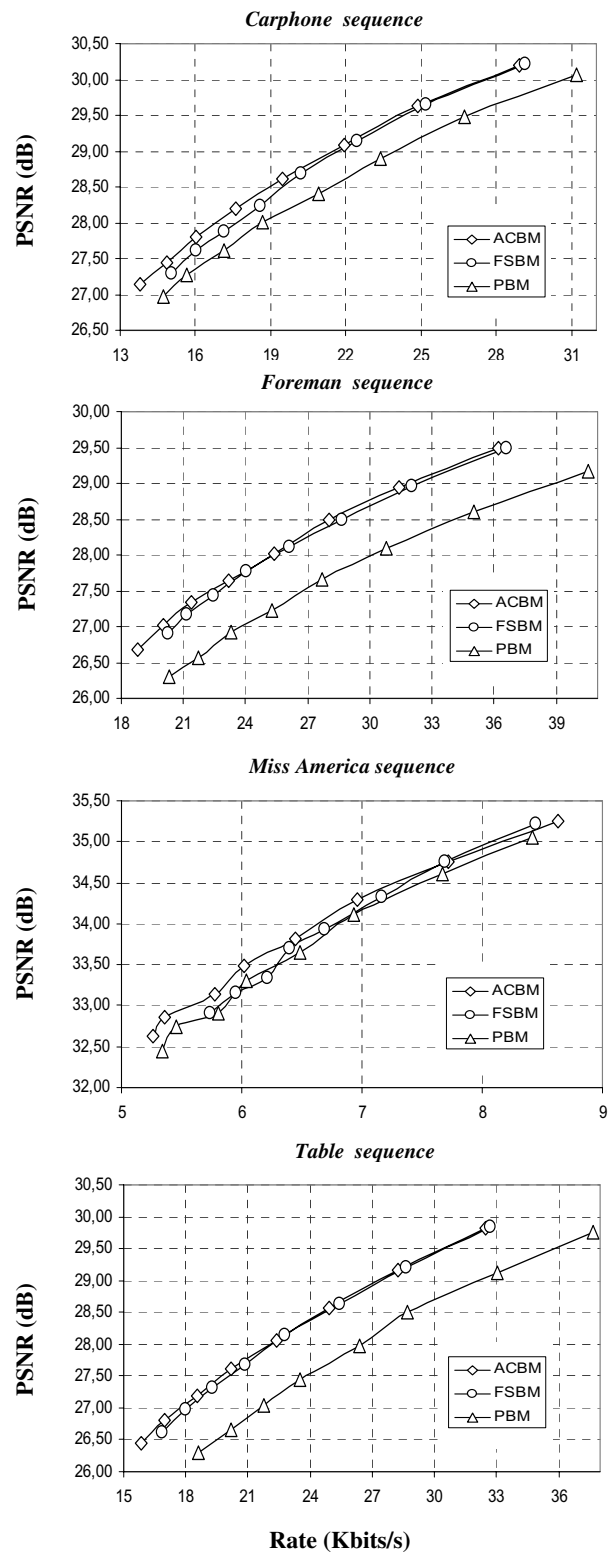

**Figure 6. Simulation results with QCIF@10 fps**